\documentclass[a4paper,fleqn]{cas-dc}
\usepackage{cas-common}
\usepackage{tabu, booktabs}
\usepackage{graphicx}
\usepackage{comment}
\usepackage{float}
\usepackage{datetime}
\usepackage{amsmath,amsfonts,amssymb,mathrsfs}
\usepackage{natbib}
\usepackage{epsfig}
\usepackage{color}
\usepackage{bm}
\usepackage{array}
\usepackage{multirow}
\usepackage{placeins}
\usepackage{caption}
\usepackage{subcaption}
\usepackage{siunitx}
\usepackage{enumitem}
\usepackage{hyperref}
\usepackage{afterpage}
\usepackage{etoolbox}
\usepackage{xcolor}
\usepackage[export]{adjustbox}
\usepackage{tabularx}
\usepackage{caption}
\usepackage{longtable}
\usepackage{enumitem}
\usepackage{booktabs}

\usepackage{makecell} % Esencial para los encabezados multilínea con \thead
\newcolumntype{C}{>{\centering\arraybackslash}X}
\newcolumntype{L}{>{\raggedright\arraybackslash}X}

\DeclareUnicodeCharacter{202F}{FIX ME!!!!}

\def\tsc#1{\csdef{#1}{\textsc{\lowercase{#1}}\xspace}}
\tsc{WGM}
\tsc{QE}

\begin{document}

\let\WriteBookmarks\relax
\def\floatpagepagefraction{1}
\def\textpagefraction{.001}

% Título del nuevo artículo
\title[mode = title]{Accelerating Multi-scale Simulations of Nuclear Components via PCYS Interpolation Tables}

% Título corto para el encabezado
\shorttitle{Título corto}    

% =============================================================================
% Autores y Afiliaciones (Formato elsarticle)
% =============================================================================
\author[1]{Fabrizio Aguzzi}
\author[1,2]{Mart\'in S. Armoa}
\author[1,2]{C\'esar I. Pairetti}
\author[1,2]{C\'esar M. Venier}
\author[3]{Alejandro E. Albanesi}

% --- AFILIACIÓN 1: IFIR (Rosario) ---
\affiliation[1]{organization={IFIR, Instituto de F\'isica de Rosario (CONICET-UNR)},
	addressline={Ocampo y Esmeralda, Predio CONICET Rosario}, 
	city={Rosario},
	postcode={2000}, 
	state={Santa Fe},
	country={Argentina}}

% --- AFILIACIÓN 3: FCEIA - UNR (Rosario) ---
\affiliation[2]{organization={Universidad Nacional de Rosario, Facultad de Ciencias Exactas, Ingenier\'ia y Agrimensura},
	addressline={Av. Pellegrini 250}, 
	city={Rosario},
	postcode={2000}, 
	state={Santa Fe},
	country={Argentina}}

% --- AFILIACIÓN 2: CIMEC (Santa Fe) ---
\affiliation[3]{organization={CIMEC, Centro de Investigaci\'on de M\'etodos Computacionales (CONICET-UNL)},
	addressline={Col. Ruta 168 s/n, Predio CONICET Santa Fe}, 
	city={Santa Fe},
	postcode={3000}, 
	state={Santa Fe},
	country={Argentina}}

\begin{abstract}
Zirconium alloy core components in nuclear reactors, such as spacer grids and fuel cladding, undergo anisotropic dimensional changes driven by coupled irradiation creep and growth. While micromechanical crystal plasticity frameworks like the Viscoplastic Self-Consistent (VPSC) formulation capture these microstructurally driven phenomena, their integration into macroscopic Finite Element Method (FEM) solvers is computationally prohibitive for engineering-scale components. To bridge this gap, this work presents a multi-scale framework implemented within the open-source FEM solver \texttt{Code\_Aster}. The developed interface uses a 5D Interpolation Table (IT) as a static material surrogate to govern instantaneous viscoplastic responses, coupled with a periodic recalibration and first-order Taylor series linearization scheme to track microstructural drift due to radiation damage without on-the-fly database updates. The predictive accuracy, numerical stability, and performance of this Polycrystal Yield Surface (PCYS) interpolation approach are benchmarked against VPSC-FEM simulations under continuous high-dose irradiation scenarios. Material-level assessments demonstrate that the linearization scheme bounds relative errors below $1\%$ for representative deformation paths, maintaining structural compatibility. Furthermore, structural simulations of a spacer grid domain revealed meaningful computational savings, overcoming the multi-scale computational penalty while preserving microstructural fidelity. The proposed framework shows potential for multiphysics
structural assessments and safety margin evaluations of core internals over operational lifespans.
\end{abstract}

% Palabras clave

\maketitle

\section{Introduction} \label{sec:intro}

Zirconium alloys are widely used as structural materials in the core of nuclear power plants, particularly for fuel cladding tubes, spacer grids, and pressure tubes, owing to their low thermal neutron absorption cross-section, adequate mechanical properties, and excellent corrosion resistance under operating conditions \citep{motta2015zirconium, griffiths2015review}. However, inside a nuclear reactor, these components are subjected to a harsh environment characterized by high temperatures, mechanical stresses, and intense neutron fields. This environment induces
non-conservative deformation phenomena, primarily irradiation creep and irradiation-induced growth, leading to significant dimensional changes and anisotropic distortion of reactor internals during reactor
operation \citep{fidleris1988irradiation, billerey2005evolution, jiang2016grid, adamson2019irradiation}. These coupled mechanisms can ultimately compromise the structural integrity and dimensional stability of the core assemblies.

The macroscopic manifestation of both irradiation growth and creep is highly anisotropic and strongly dictated by the initial and evolving crystallographic texture of the material \citep{holt2008reactor, gicquel2023polycrystalline}. Consequently, standard isotropic or phenomenological macroscopic constitutive laws often fail to accurately predict long-term reactor component behavior under complex, multiaxial stress states. In this context, micromechanically based crystal plasticity frameworks have been developed. Over the past decades, the Viscoplastic Self-Consistent (VPSC) formulation \citep{molinari1987self, lebensohn1993self} has emerged as a robust tool to simulate the representative volume element (RVE) of zirconium polycrystals, successfully accounting for grain interactions, slip system activities, and microstructural evolution under irradiation \citep{christodoulou1996modeling, patra2017crystal}.

Despite their physical accuracy, embedding polycrystal models directly into full-scale structural analyses via macro-scale Finite Element Method (FEM) implies some implementation challenges. In a direct concurrent multiscale approach, the VPSC algorithm must be solved iteratively at every single integration point and time increment of the FE mesh. For complex three-dimensional components, such as fuel assembly components or cladding tubes under non-uniform fluxes, this scheme renders engineering-scale simulations computationally prohibitive \citep{galan2014improved, aguzzi2025toolbox,aguzzi2026multiphysics}. Indeed, even when operating strictly at the macroscopic engineering scale, performing full 3D FEM structural assessments over long operational times becomes restrictive when evaluating a large number of core components or executing probabilistic safety margin analyses \citep{pandey2018understanding, prabhu2020surrogate}. This severe computational cost has driven the nuclear engineering community to actively seek alternative accelerated representations and efficient numerical strategies to manage multi-physical or multi-scale interactions in reactor cores \citep{prabhu2020surrogate}.

Database-driven and surrogate modeling strategies offer a path to reconcile microstructural fidelity with computational feasibility. At the macroscopic component level, surrogates have been developed to replace demanding 3D FE representations for predicting complex in-reactor dimensional changes, such as the contact time and gap sag in irradiated fuel channels \citep{pandey2018understanding, prabhu2020surrogate}. Conversely, at the material scale, a widely adopted method consists of pre-calculating the Polycrystal Yield Surface (PCYS) to build Interpolation Tables (IT) or databases that map the instantaneous stress-strain rate responses in a reduced deviatoric space \citep{mcginty2001multiscale, tome2023material}. While these static, material-level interpolation databases accelerate local numerical evaluations, they implicitly assume a frozen microstructural state—meaning that variations in crystallographic texture or Critical Resolved Shear Stress (CRSS) driven by accumulating radiation damage (dpa) are neglected over time. While some frameworks achieve coupling via complex analytical approximations \citep{brenner2002quasi, gicquel2023polycrystalline}, a flexible, low-cost numerical mechanism capable of rectifying the microstructural drift of static databases under long-term irradiation scenarios remains desirable.

In this work, we propose a multi-scale framework that bridges this gap by introducing a static interpolation table coupled with a periodic recalibration strategy implemented in the open-source FEM solver Code\_Aster \citep{ASTER}. The proposed interface applies a 5D Interpolation Table to govern the instantaneous viscoplastic response, while concurrently implementing a periodic recalibration and first-order Taylor series linearization scheme. This scheme tracks the drift and
updates of the macroscopic compliance ($\bar{M}$) and irradiation growth ($\bar{\dot{\varepsilon}}^0$) tensors with minimal VPSC solver calls. The main objective of this study is to evaluate the performance, accuracy, and speed-up gains of this IT-linearization methodology against VPSC-FEM concurrent simulations under representative nuclear application scenarios.

The remainder of this paper is organized as follows. Section \ref{sec:methodology} outlines the multi-scale computational framework, the material models, and the accelerated IT linearization algorithm. Section \ref{sec:resultados} presents the numerical benchmarks alongside a comparative analysis of precision and computational cost at both material and component scales. Finally, Section \ref{sec:conclusions} summarizes the main conclusions and future perspectives of this research. For reference, a comprehensive list of symbols and acronyms is provided in the Nomenclature section at the end of the manuscript.

\section{Computational Framework and Multiscale Strategy}\label{sec:methodology}

The multi-scale strategy developed in this work bridges the gaps between single-crystal physical mechanisms, meso-scale polycrystal interactions, and macro-scale finite element simulations. This section outlines the overarching boundary value problem governing the continuum component, details the underlying crystallographic constitutive laws, and presents the formulation of the high-performance surrogate interpolation scheme along with its algorithmic integration into the FE solver.

\subsection{Numerical Formulation of the Macro-Scale Boundary Value Problem}
\label{subsec:macro_formulation}

To resolve the non-linear, quasi-static structural mechanics problem at the nuclear component level, the finite element solver must integrate a tightly coupled system of governing equations representing conservation laws, local material responses, and history-dependent microstructural updates \citep{agouzal2024reduced}. Over the considered continuum domain, this initial-boundary value problem is mathematically formulated as:

\begin{equation}
\left\{
\begin{aligned}
    -\boldsymbol{\nabla} \cdot \boldsymbol{\sigma} &= \mathbf{f} \\  
    \boldsymbol{\sigma} &= \mathcal{F}^{\sigma}(\boldsymbol{\varepsilon}_{\text{FE}}, \boldsymbol{\beta}) \quad \text{subject to } \text{BCs} \\
    \dot{\boldsymbol{\beta}} &= \mathcal{F}^{\beta}(\boldsymbol{\sigma}, \boldsymbol{\beta})
\end{aligned}
\right.
\end{equation}

\noindent where $\mathbf{f}$ represents the body force vector, $\boldsymbol{\sigma}$ denotes the macroscopic Cauchy stress tensor field, and $\boldsymbol{\varepsilon}_{\text{FE}}$ is the total strain tensor derived from the symmetric gradient of the displacement field $\mathbf{u}$, defined as $\boldsymbol{\varepsilon}_{\text{FE}} = \boldsymbol{\nabla}_s \mathbf{u} = \frac{1}{2}(\boldsymbol{\nabla} \mathbf{u} + \boldsymbol{\nabla}^T \mathbf{u})$. The history-dependent state of the material is tracked via the internal variable vector $\boldsymbol{\beta}$, which encapsulates the microstructural attributes of the polycrystal, including crystallographic orientation (texture), grain morphology, and the current hardening state of the active slip systems within individual grains.

The system of equations encapsulates three interconnected physical and mathematical definitions. The first equation enforces momentum balance and mechanical equilibrium across the continuum. The second equation introduces the non-linear constitutive operator $\mathcal{F}^{\sigma}$, mapping the local stress state as a joint function of the mechanical strain and the instantaneous microstructural state. Finally, the third equation governs the kinetic evolution of the internal variables through the differential operator $\mathcal{F}^{\beta}$, capturing the microstructural drift induced by irradiation and plastic deformation over time. Within our multiscale framework, this evolution is efficiently regularized via the PCYS interpolation extrapolation strategy detailed in the following sections.

\subsection{Crystallographic and Meso-Scale Constitutive Modeling}
\label{subsec:crystal_meso_model}

\subsubsection{Irradiation-Induced Growth Mechanisms}
Following the reaction-diffusion formulation proposed by \cite{patra2017crystal}, we describe irradiation growth as a non-conservative deformation process at the single-crystal (microscopic) level. The growth strain-rate tensor, $\dot{\varepsilon}_{kl}^{(\text{growth})}$, is calculated through a linear superposition of the contributions along the principal crystallographic directions $j \in \{\mathbf{a}_1,\mathbf{a}_2,\mathbf{a}_3,\mathbf{c}\}$ of the hexagonal close-packed (HCP) lattice, where $\mathbf{a}_1$, $\mathbf{a}_2$, and $\mathbf{a}_3$ represent the coplanar axes in the basal plane, and $\mathbf{c}$ denotes the orthogonal prism axis. This tensor is expressed as:
\begin{equation}\label{eq:growth_tensor}
	\dot{\varepsilon}_{kl}^{(\text{growth})} = \sum_{j} \dot{\varepsilon}_{\text{growth}}^j (b_k^j b_l^j), \quad k,l \in \{x,y,z\}
\end{equation}
In this expression, $b_k^j$ and $b_l^j$ represent the Cartesian components of the normalized Burgers vector $\hat{\mathbf{b}}^j$ projected onto the macroscopic sample reference frame (i.e., $b_k^j = \hat{\mathbf{b}}^j \cdot \mathbf{e}_k$, where $\{\mathbf{e}_x, \mathbf{e}_y, \mathbf{e}_z\}$ define the sample axes). The scalar rate $\dot{\varepsilon}_{\text{growth}}^j$ accounts for the combined effects of point defect absorption:
\begin{equation}\label{eq:growth_rates}
	\dot{\varepsilon}_{\text{growth}}^j = \dot{\varepsilon}_{\text{climb}}^j + \dot{\varepsilon}_{\text{GB}}^j
\end{equation}
where $\dot{\varepsilon}_{\text{climb}}^j$ and $\dot{\varepsilon}_{\text{GB}}^j$ represent the local strain rates associated with dislocation climb and grain boundary absorption, respectively.

\subsubsection{Crystallographic Model for Irradiation Creep}
In addition to growth, irradiation-induced creep is modeled as a stress-dependent relaxation mechanism at the single-crystal level. This phenomenon, which is intrinsically linked to growth due to internal stress incompatibilities between neighboring grains, is assumed to be proportional to the radiation dose rate $\dot{\phi}$. 

The kinematics of single-crystal deformation dictate the evolution of the local microscopic creep strain-rate tensor, $\dot{\varepsilon}_{kl}^{(\text{creep})}$. This tensor is obtained by summing the shear rates $\dot{\gamma}_{\text{creep}}^{j}$ across all active crystallographic slip systems $j$:
\begin{equation}\label{eq:kinematic_connection}
    \dot{\varepsilon}_{kl}^{(\text{creep})} = \sum_{j} m_{kl}^j \dot{\gamma}_{\text{creep}}^{j}
\end{equation}
where $m_{kl}^j = \frac{1}{2} (n_k^j b_l^j + n_l^j b_k^j)$ is the symmetric Schmid tensor for system $j$, defined by its slip plane normal $\mathbf{n}^j$ and slip direction $\mathbf{b}^j$. Following \cite{patra2017crystal}, the shear rate on each system is governed by a linear phenomenological law:
\begin{equation}\label{eq:creep_shear}
	\dot{\gamma}_{\text{creep}}^{j} = B \left( \frac{\rho_d^j}{\rho_\text{ref}} \right) \tau^j \dot{\phi}
\end{equation}
Here, $B$ denotes the crystallographic irradiation creep compliance, $\tau^j = m_{kl}^j \sigma_{kl}$ is the resolved shear stress acting on the system, and $\rho_d^j/\rho_\text{ref}$ represents the normalized dislocation density on the active slip system. 

Consequently, by utilizing a proper homogenization scheme over the crystal domains, the macroscopic creep behavior exhibits a linear relation with the macroscopic stress state. For an isotropic polycrystal,
or under a simplified macro-scale approximation, this simplifies to a second-order tensor relationship:
\begin{equation}\label{eq:macro_creep}
	\dot{\bar{\varepsilon}}_{ij}^{\text{creep}} = B_{0} \bar{\sigma}_{ij} \dot{\phi}
\end{equation}
where $\dot{\bar{\varepsilon}}_{ij}^{\text{creep}}$ and $\bar{\sigma}_{ij}$ are the components of the macroscopic strain-rate and stress tensors, respectively, and $B_0$ is the macroscopic compliance coefficient. Parameter values and additional details regarding the integration of these single-crystal constitutive laws into the meso-scale framework are detailed in Appendix \ref{apendiceA}.

\subsubsection{Viscoplastic Self-Consistent Model (VPSC)}
\label{VPSC}
The mechanical response of the Zircaloy-2 polycrystal is simulated using the VPSC formulation \citep{molinari1987self, tome1993self,patra2017crystal}. In this meso-scale framework, the total microscopic strain rate at the grain level, $\dot{\varepsilon}_{ij}$, is defined by the purely viscoplastic superposition of irradiation-induced growth and creep:
\begin{equation}
\dot{\varepsilon}_{ij} = \dot{\varepsilon}_{ij}^{\text{vp}} = \dot{\varepsilon}_{ij}^{(\text{growth})} + \dot{\varepsilon}_{ij}^{(\text{creep})}
\end{equation}
To determine the response of the effective medium (polycrystal), a tangent linearization scheme is adopted \citep{tome2023material}, yielding the polycrystal fourth-order constitutive relationship for the Homogeneous Effective Medium (HEM):
\begin{equation}\label{eq:vpsc_macro}
\bar{\dot{\varepsilon}}_{ij} = \bar{\dot{\varepsilon}}_{ij}^{\text{vp}} = \bar{M}_{ijkl} \bar{\sigma}_{kl} + \bar{\dot{\varepsilon}}_{ij}^0
\end{equation}
where $\bar{\dot{\varepsilon}}_{ij}$ (or $\bar{\dot{\varepsilon}}_{ij}^{\text{vp}}$) and $\bar{\sigma}_{kl}$ represent the volume-averaged polycrystal viscoplastic strain rate and stress tensors, respectively, while $\bar{M}_{ijkl}$ denotes the fourth-order polycrystal compliance tensor. 

The term $\bar{\dot{\varepsilon}}_{ij}^0$ represents the polycrystal back-extrapolated growth strain rate. Mathematically, it acts as a pseudo-residual or intercept tensor arising from the first-order Taylor expansion utilized in the tangent linearization of the non-linear polycrystal response. Physically, this tensor accounts for the stress-independent dimensional changes driven by anisotropic irradiation growth, combined with the baseline microstructural offset required to scale the grain-level viscoplastic interactions up to the homogeneous effective medium scale.

To fulfill the self-consistent requirement, the framework treats each grain orientation as an ellipsoidal inclusion embedded within the HEM. The localized stress $\sigma_{kl}^c$ and strain rate $\dot{\varepsilon}_{ij}^c$ fields of each distinct grain $c$ are coupled to the macroscopic fields through an Eshelby-type interaction tensor $\tilde{M}_{ijkl}$, such that:
\begin{equation}
    (\dot{\varepsilon}_{ij}^c - \bar{\dot{\varepsilon}}_{ij}) = -\tilde{M}_{ijkl} (\sigma_{kl}^c - \bar{\sigma}_{kl})
\end{equation}
where $\tilde{M}_{ijkl}$ depends on the grain morphology and the macroscopic compliance $\bar{M}_{ijkl}$. Rather than detailing the extensive system of iterative equations widely established in literature, the operational logic of the self-consistent loop used to achieve the macroscopic volume averages ($\bar{\sigma}_{kl} = \langle \sigma_{kl}^c \rangle$ and $\bar{\dot{\varepsilon}}_{ij} = \langle \dot{\varepsilon}_{ij}^c \rangle$) can be consulted via the comprehensive algorithmic flowchart detailed in \cite{tome2023material}.

\subsection{Surrogate Model Generation: Interpolation Tables (IT)}
\label{subsec:surrogate_IT}

To circumvent the prohibitive computational cost that arises when embedding detailed crystal-plasticity models directly into macroscopic Finite Element Method (FEM) solvers, the non-linear polycrystal response is mapped into a high-performance surrogate database. Although standalone Viscoplastic Self-Consistent (VPSC) formulations efficiently solve a representative volume element, executing it concurrently at every single integration point and time increment of an FE mesh renders large-scale engineering component simulations computationally restrictive. To reduce this numerical cost, the VPSC model is utilized as an offline pre-processor to construct an Interpolation Table (IT) representing the Polycrystal Yield Surface (PCYS).

This database provides an efficient mapping within a 5D deviatoric stress space: entering with the current stress state $\boldsymbol{\sigma}$ to instantly retrieve the viscoplastic strain rate $\boldsymbol{\dot{\varepsilon}}^{\text{IT}}$, or vice versa, thus
bypassing the thousands of internal iterations required by a
direct self-consistent solve. The mathematical foundation of the 5D angular discretization, the specialized probing strategy, and the optimization routines required to build this discrete database are comprehensively detailed in Appendix \ref{apendiceB}.

Crucially, since the database is constructed offline for a fixed reference state, it implicitly assumes that the underlying crystallographic texture and the Critical Resolved Shear Stress (CRSS) of the slip systems remain stationary. To overcome this limitation and account for continuous microstructural evolution during irradiation without requiring computationally intensive on-the-fly table updates, a PCYS interpolation approach is introduced. This strategy dynamically adjusts the surrogate outputs by coupling the static IT response with a macro-scale Taylor extrapolation framework, as presented below.

\subsubsection{First-Order Extrapolation Strategy for Microstructural Drift}
\label{subsec:extrapolation_strategy}
To minimize computational cost while capturing history-dependent phenomena, the evolution of the macroscopic viscoplastic compliance $\bar{M}$ and the back-extrapolated growth rate $\bar{\dot{\varepsilon}}^0$ at each integration point is predicted using a periodic recalibration logic. This PCYS interpolation strategy depends on a user-defined refresh interval $N$ (typically $N \in \{4, 6, 8, 10\}$ steps) and operates through three distinct execution phases, conceptually illustrated in Fig. \ref{fig:vpsc_linearization}:

\begin{enumerate}
    \item \textbf{Recalibration and Error Monitoring ($i_{\text{step}} \pmod N = 0$):} Every $N$ steps, the surrogate database is bypassed and a full VPSC cell computation is executed. This calculates the exact current tensors, directly updating the microstructural state variables. 
    
\item \textbf{Evolution Rate Estimation ($i_{\text{step}} \pmod N = 1$):} Immediately following a recalibration step, a single reference VPSC call is executed in the subsequent increment to compute the exact numerical rates of change (slopes):
    \begin{equation}
        \dot{\bar{M}} = \frac{\bar{M}_{\text{ref2}} - \bar{M}_{\text{ref1}}}{\Delta t}, \quad \ddot{\bar{\varepsilon}}^0 = \frac{\bar{\dot{\varepsilon}}^0_{\text{ref2}} - \bar{\dot{\varepsilon}}^0_{\text{ref1}}}{\Delta t}
    \end{equation}
    where $\Delta t$ denotes the simulation time increment. These numerical slopes capture the localized evolution of hardening and irradiation growth along the current deformation path.

\item \textbf{Linear Tangent Approximation (Remaining steps):} For all intermediate steps between recalibrations, full meso-scale
calculations are bypassed. The evolution of the tensors is predicted via a first-order Taylor expansion using the latest available rates:
    \begin{equation}\label{eq:taylor_extrap}
        \bar{M}_{\text{tan}} \approx \bar{M}_{\text{ref2}} + \dot{\bar{M}} \Delta t, \quad \bar{\dot{\varepsilon}}^0_{\text{tan}} \approx \bar{\dot{\varepsilon}}^0_{\text{ref2}} + \ddot{\bar{\varepsilon}}^0 \Delta t
    \end{equation}
    Consequently, the extrapolated macroscopic viscoplastic strain rate $\boldsymbol{\dot{\varepsilon}}^{\text{vp}}$ to be used in the mechanical solver is evaluated as:
    \begin{equation}\label{eq:evp_extrap}
        \boldsymbol{\dot{\varepsilon}}^{\text{vp}} = \bar{M}_{\text{tan}} \boldsymbol{\sigma}^ + \bar{\dot{\varepsilon}}^0_{\text{tan}}
    \end{equation}
\end{enumerate}

\begin{center}
    \centering
    \includegraphics[width=0.45\textwidth]{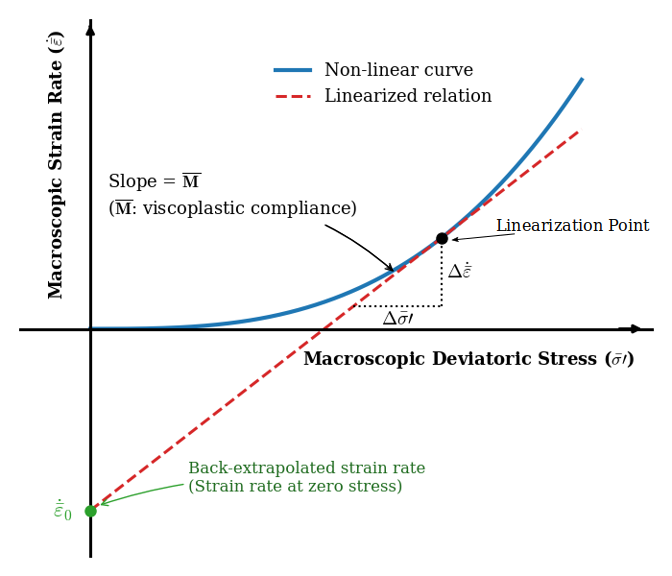} 
    \captionof{figure}{Schematic representation of the VPSC non-linear constitutive response and its local linearization. The slope of the linearized relation defines the macroscopic viscoplastic compliance ($\bar{M}$), while the intercept at zero stress yields the back-extrapolated strain rate ($\bar{\dot{\varepsilon}}_0$).}
    \label{fig:vpsc_linearization}
\end{center}

\subsubsection{VPSC-CAFEM Coupling Architecture via Code\_Aster}
\label{Acople VPSC-FE}
The multiscale interface integrates the localized, microstructurally driven constitutive response governed by the extrapolation strategy (Sect. \ref{subsec:extrapolation_strategy}) into the implicit continuum FEM solver \texttt{Code\_Aster}. At this point, the different
kinematic assumptions of both solvers must be distinguished. The standalone viscoplastic self-consistent formulation (VPSC-SA) operates under a rigid-viscoplastic assumption, directly mapping macroscopic deviatoric stresses to viscoplastic strain rates without evaluating elastic compliance. To overcome this limitation and support structural analysis under complex boundary conditions, the coupling framework incorporates macroscopic elasticity at the integration point level. 

Under a small-strain additive decomposition regime, the total continuum strain increment $\Delta \boldsymbol{\varepsilon}$ is decoupled into macroscopic elastic ($\Delta \boldsymbol{\varepsilon}^{\text{e}}$) and viscoplastic ($\Delta \boldsymbol{\varepsilon}^{\text{vp}}$) contributions:
\begin{equation}
\Delta\boldsymbol{\varepsilon} = \Delta \boldsymbol{\varepsilon}^{\text{e}} + \Delta \boldsymbol{\varepsilon}^{\text{vp}} = \mathbf{C}^{-1}:\Delta\boldsymbol{\sigma} + \Delta \boldsymbol{\varepsilon}^{\text{vp}}
\end{equation}
\noindent where $\mathbf{C}$ represents the self-consistent elastic stiffness tensor of the polycrystal—evaluated at the beginning of the time increment—and $\Delta\boldsymbol{\sigma}$ denotes the Cauchy stress increment. While $\Delta \boldsymbol{\varepsilon}^{\text{e}}$ depends directly on the current stress increment, the viscoplastic strain increment $\Delta\boldsymbol{\varepsilon}^{\text{vp}}$ is a function of the instantaneous stress state and the history-dependent evolution of the internal variable vector $\boldsymbol{\beta}$.

To maintain reference frame consistency with the anisotropic constitutive relations, the global strain increment $\Delta\boldsymbol{\varepsilon}_{\text{FE}}$ and the step time $\Delta t$ provided by \texttt{Code\_Aster} are mapped onto the local material coordinate system ($^*$) via the second-order rotation matrix $\boldsymbol{R}$:
\begin{equation}
    \Delta\boldsymbol{\varepsilon}^* = \boldsymbol{R} \, \Delta\boldsymbol{\varepsilon}_{\text{FE}} \, \boldsymbol{R}^T, \quad \boldsymbol{\sigma}^* = \boldsymbol{R} \, \boldsymbol{\sigma} \, \boldsymbol{R}^T
\end{equation}
Assuming a quasi-static regime over the interval, the local orientation matrix $\boldsymbol{R}$ is assumed to remain stationary within the corotational frame.

The local stress state at the end of the step, $\boldsymbol{\sigma}^{t+\Delta t^*}$, is computed by considering an initial purely elastic trial state subsequently relaxed by viscoplastic flow:
\begin{equation}
    \boldsymbol{\sigma}^{t+\Delta t^*} = \boldsymbol{\sigma}^{t^*} + \mathbf{C}:\left(\Delta\boldsymbol{\varepsilon}^* - \Delta\boldsymbol{\varepsilon}^{\text{vp}*} \right)
\end{equation}
By replacing the increment $\Delta\boldsymbol{\varepsilon}^{\text{vp}*}$ with the instantaneous rate integrated over the time step ($\boldsymbol{\dot{\varepsilon}}^{\text{vp}*} \Delta t$), a local Newton-Raphson (NR) iterative scheme is formulated to find the stress increment $\Delta\boldsymbol{\sigma}^*$ that satisfies structural compatibility. The corresponding residual tensor field $\mathbf{X}\left( \Delta\boldsymbol{\sigma}^* \right)$ is defined as:
\begin{equation}
    \mathbf{X}\left( \Delta\boldsymbol{\sigma}^* \right) = \mathbf{C}^{-1}:\Delta\boldsymbol{\sigma}^* + \boldsymbol{\dot{\varepsilon}}^{\text{vp}*}(\boldsymbol{\sigma}^*) \Delta t - \Delta\boldsymbol{\varepsilon}^*_{\text{FE}}
\end{equation}
where the viscoplastic strain rate $\boldsymbol{\dot{\varepsilon}}^{\text{vp}*}$ is evaluated using the extrapolated constitutive relation defined in Eq. \ref{eq:evp_extrap}. If the residual field violates the convergence tolerance at iteration $k$, the local stress increment is updated for the subsequent iteration $k+1$ via:
\begin{equation}\label{ec:correcciondeTension}
    \left( \Delta\boldsymbol{\sigma}^*\right)_{k+1} = \left( \Delta\boldsymbol{\sigma}^*\right)_{k} - \mathbf{J}^{*-1}_{\text{NR}}\left((\Delta\boldsymbol{\sigma}^*)_{k}\right) : \mathbf{X}\left((\Delta\boldsymbol{\sigma}^*)_{k}\right)
\end{equation}

The non-linear search is regularized by the local Jacobian matrix $\mathbf{J}^*_{\text{NR}}$, which explicitly accounts for both elastic compliance and the extrapolated viscoplastic tangent response:
\begin{equation}
    \mathbf{J}^*_{\text{NR}} = \frac{\partial\mathbf{X}(\Delta\boldsymbol{\sigma}^*)}{\partial(\Delta\boldsymbol{\sigma}^*)} = \mathbf{C}^{-1} + \bar{M}_{\text{tan}} \Delta t
\end{equation}

To prevent convergence distortions in loading paths where certain strain components approach zero, a component-wise error metric $\chi$ is adopted \citep{mcginty2001multiscale}. This metric normalizes local residuals using the maximum component of the input strain increment:
\begin{equation}
    \chi = \sqrt{\sum_i \sum_j \left(\frac{|\Delta\varepsilon_{\text{FE}}^{ij}|}{\max(|\Delta\varepsilon_{\text{FE}}^{ij}|)} \mathbf{X}^{ij} \right)^2}
\end{equation}

Once convergence ($\chi < \text{TOL}_{\text{local}}$) is achieved, the updated local Cauchy stress $\boldsymbol{\sigma}^{t+\Delta t^*}$ and the consistent tangent operator $\mathbf{C}^{\text{tg*}} = \mathbf{J}^{*-1}_{\text{NR}}$ must be rotated back to the global reference frame to preserve compatibility with the global equilibrium iterations of \texttt{Code\_Aster}:
\begin{equation}
    \boldsymbol{\sigma}^{t+\Delta t} = \boldsymbol{R}^T \boldsymbol{\sigma}^{t+\Delta t^*} \boldsymbol{R}, \quad \mathbf{C}^{\text{tg}} = \mathcal{R} \, \mathbf{C}^{\text{tg*}} \, \mathcal{R}^T
\end{equation}
where $\mathcal{R}$ is the fourth-order rotation operator, which maps the components explicitly according to:
\begin{equation}
    C^{\text{tg}}_{ijkl} = R_{im} R_{jn} R_{ko} R_{lp} \, C^{\text{tg*}}_{mnop}
\end{equation}

\section{Results} \label{sec:resultados}

In this section, the performance, accuracy, and computational efficiency of the proposed PCYS interpolation acceleration strategy are evaluated through numerical benchmarks. First, a material-level assessment is carried out on a single integration point to investigate the sensitivity of the Taylor-series linearization scheme to the refresh interval $N$. Second, the framework is validated in a representative structural application using the finite element solver Code\_Aster, evaluating the spatial and temporal evolution of macroscopic fields. Finally, a quantitative assessment of the computational speed-up gains is presented.

\subsection{Material-Level Performance and Drift Sensitivity Analysis}
\label{subsec:material_validation}

The first-order Taylor linearization strategy described in Section \ref{sec:methodology} is tested under a continuous irradiation regime up to a high damage dose to evaluate the numerical drift induced by bypassing the VPSC solver. The reference solution is obtained by a direct concurrent coupling where the exact macroscopic compliance ($\bar{M}$) and growth rate ($\bar{\dot{\varepsilon}}^0$) tensors are computed via VPSC at every single time increment ($\Delta t$). The complete set of single-crystal constitutive parameters, elastic constants, and irradiation properties utilized in these simulations are detailed in Appendix \ref{apendiceA}. Crucially, these material properties and calibration parameters remain identical and are uniformly applied across all numerical benchmarks and component-scale simulations presented throughout this work.

\begin{figure*}[t]
    \centering
    \includegraphics[width=0.95\textwidth]{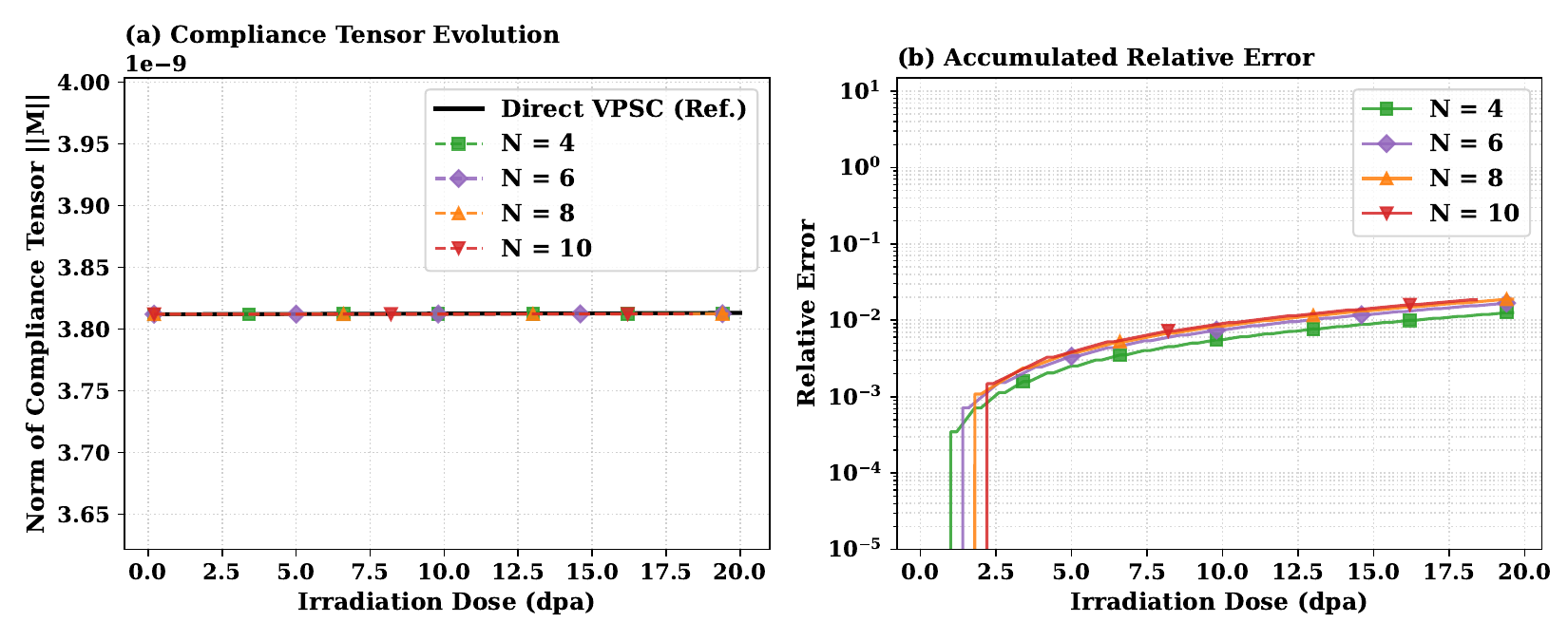}
    \caption{Standalone material-point sensitivity analysis of the microstructural linearization scheme for different refresh intervals $N$under homogeneous
standalone conditions (without FEM coupling): (a) compliance tensor evolution, and (b) accumulated relative error as a function of irradiation dose.}
    \label{fig:material_drift_comparison}
\end{figure*}

Fig. \ref{fig:material_drift_comparison} illustrates the comparative evolution of the norm of the compliance tensor $\bar{M}$ and the macroscopic growth components as a function of the irradiation dose (dpa) for different user-defined refresh intervals ($N \in \{4, 6, 8, 10\}$). 

\begin{figure*}[t!]
    \nopagebreak
    \includegraphics[width=\textwidth]{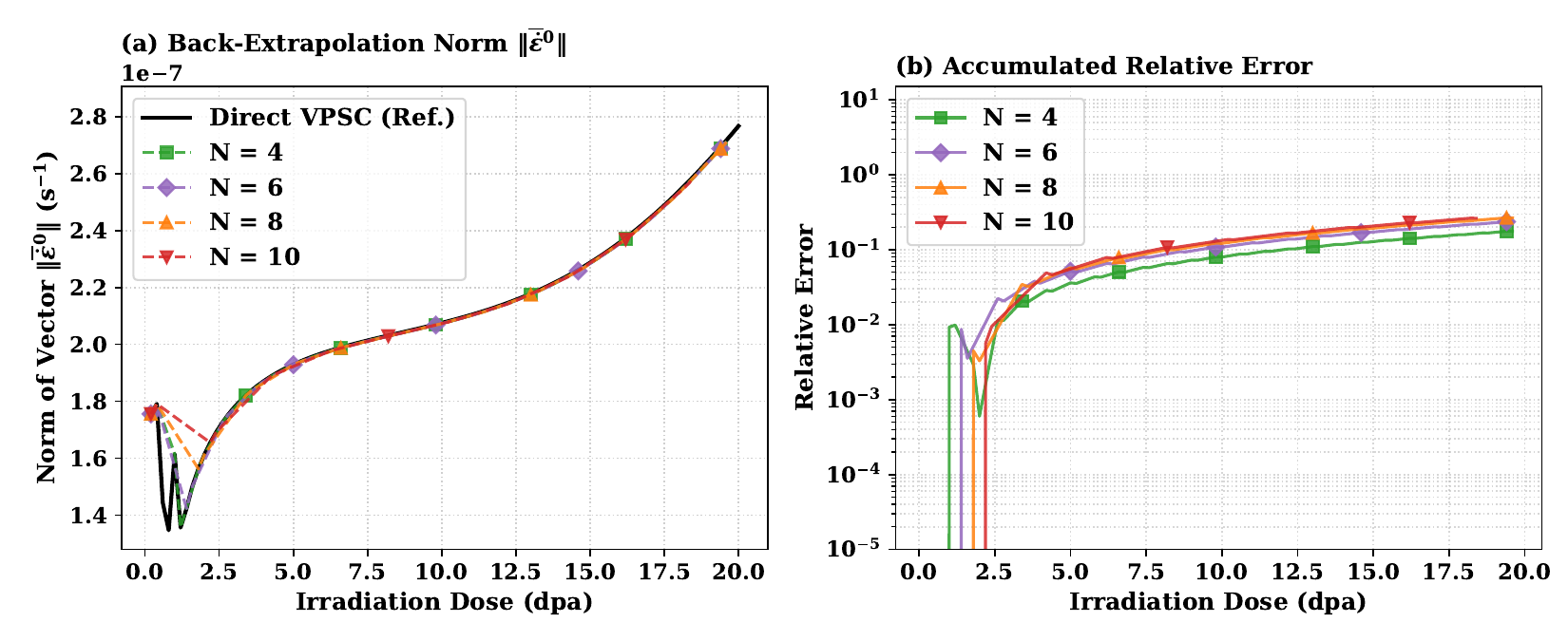}
    \caption{Standalone material-point sensitivity and numerical drift analysis of the back-linearization strain rate vector $\dot{\bar{\varepsilon}}^0$ for different user-defined refresh intervals $N$ (no FEM coupling): (a) evolution of the scalar norm $\|\dot{\bar{\varepsilon}}^0\|$ as a function of the irradiation dose, and (b) its corresponding accumulated relative error showing the characteristic sawtooth truncation pattern driven by the periodic recalibration routine.}
    \label{fig:dbar0_drift_comparison}
\end{figure*}

In addition to the compliance tensor, the numerical stability of the sub-stepping approach is linked to the calculation of the back-linearization strain rate tensor, $\bar{\dot{\varepsilon}}^0$, which represents the stress-free growth and shape-change rate induced by microstructural evolution under irradiation. Fig.~\ref{fig:dbar0_drift_comparison} displays the comparative sensitivity analysis for the norm $\|\bar{\dot{\varepsilon}}^0\|$ and its corresponding relative accumulated error.

As observed in Fig.~\ref{fig:dbar0_drift_comparison}a, the trajectory of $\|\bar{\dot{\varepsilon}}^0\|$ exhibits a non-linear transient behavior during the early stages of irradiation ($\le 2.5~\text{dpa}$), where sharp texture modification and rapid initial hardening occur. For conservative intervals ($N = 4$), the Taylor-expansion approximation tracks the reference curve, keeping the mathematical drift strictly under $0.1\%$. 

For larger recalibration intervals ($N = 10$), localized spikes in the relative error are observable during the periods of maximum non-linearity, reaching values close to $5\%$. Nevertheless, as depicted by the distinctive sawtooth pattern in Fig.~\ref{fig:dbar0_drift_comparison}b, the discrete nature of the periodic recalibration routine ($i_{\text{step}} \bmod N = 0$) abruptly truncates error propagation at the end of each interval. The algorithm successfully resets $\bar{\dot{\varepsilon}}^0$ to its exact VPSC solution, forcing the error down to negligible levels before any numerical divergence can compromise the global Newton-Raphson material subroutine. Once the microstructure stabilizes at higher doses ($> 5~\text{dpa}$), the accumulated error for all configurations remains safely bounded well below the $1\%$ threshold.

\begin{figure*}
    \nopagebreak
    \includegraphics[width=\textwidth]{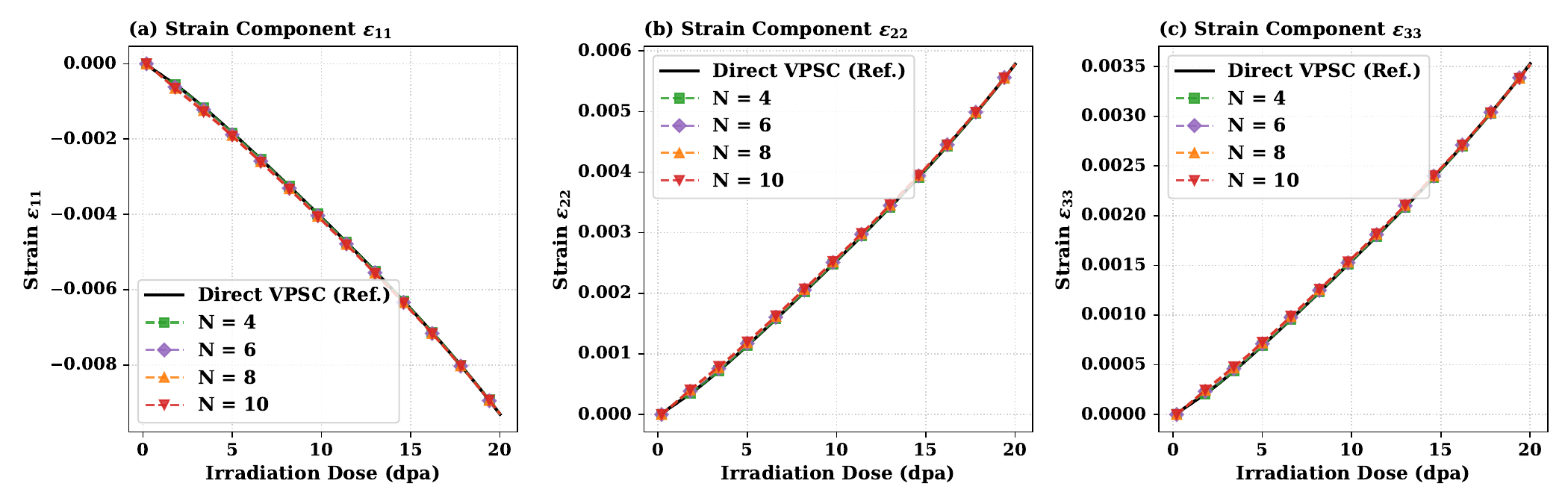}
    \caption{Comparison of polycrystal visco-plastic strain evolution between the full direct-coupling VPSC reference solution and different refresh intervals ($N = 4, 6, 8, 10$) evaluated at a standalone material point (no FEM coupling). Panels show the independent strain components corresponding to the principal directions of the nuclear spacer grid texture: (a) Normal component ($\varepsilon_{11}$), (b) Transverse component ($\varepsilon_{22}$), and (c) Rolling component ($\varepsilon_{33}$) as a function of the accumulated irradiation dose.}
    \label{fig:strain_components_comparison}
\end{figure*}

As expected, increasing the linearization interval $N$ directly reduces the number of full VPSC calls, but accumulates a numerical drift due to the linear approximation of the microstructural trajectory. To evaluate the polycrystal kinematic consequence of the sub-stepping scheme, Fig. \ref{fig:strain_components_comparison} presents the evolution of the independent strain components as a function of the irradiation dose. For these simulations, the reduced texture of 13 grains discussed in \citep{aguzzi2025toolbox} was utilized. Given the symmetry and processing history of the component under study, these axes correspond to the Normal ($\varepsilon_{11}$), Transverse ($\varepsilon_{22}$), and Rolling ($\varepsilon_{33}$) directions of the zirconium-alloy nuclear fuel spacer grid. Consistent with the compliance tensor trends, the strain paths calculated for low refresh intervals ($N = 4$) display agreement with the direct coupling VPSC reference solution across all three coordinate axes. For larger intervals ($N = 6$ and $N = 8$), slight deviations accumulate during early irradiation stages, where rapid microstructure evolution takes place; however, the drift remains bounded. Even for the $N = 10$ configuration, the kinematic deviations are contained by the periodic recalibration steps. This demonstrates that minor localized drifts in the polycrystal compliance tensor do not trigger severe macroscopic strain deviations, thereby ensuring the numerical stability required for large-scale engineering finite element simulations.

\FloatBarrier

To further quantify these discrepancies, the continuous evolution of the macroscopic strain tensor relative error norm is evaluated. To prevent horizontal overflows within the standard double-column layout, this error metric is mathematically defined using a compact notation as follows:
\begin{equation}
\parallel \mathbf{e} \parallel_2 [\%] = \frac{\left[ \sum_{i=1}^3 (\varepsilon_{ii} - \varepsilon_{ii,\text{ref}})^2 \right]^{1/2}}{\left[ \sum_{i=1}^3 \varepsilon_{ii,\text{ref}}^2 \right]^{1/2}} \times 100
\end{equation}
where $\varepsilon_{ii}$ and $\varepsilon_{ii,\text{Ref}}$ represent the independent strain components obtained with the accelerated multi-dimensional interpolation table (IT) framework and the standard standalone direct VPSC reference solution, respectively, at each irradiation increment.

As plotted in Fig. \ref{fig:strain_error_evolution}, during the early incubation and transient hardening stage (under $3~\text{dpa}$), the error norm exhibits a localized peak for all configurations, reaching approximately $17.5\%$ for $N = 10$. This transient drift is driven by the non-linear nature of the initial microstructural evolution, where a first-order Taylor series linearization temporarily underpredicts the rate changes. Crucially, as the microstructural evolution stabilizes at higher irradiation doses ($> 7.5~\text{dpa}$), the self-correcting mechanism of the periodic VPSC recalibrations induces asymptotic convergence. By the end of the irradiation regime, the cumulative error decays to a relative value below $2.5\%$ even for the most restrictive $N = 10$ case, confirming that the framework does not introduce systemic error propagation over extended exposure periods.

From an engineering design perspective, this error profile is highly acceptable for in-reactor component simulations. While the transient peak under $3~\text{dpa}$ appears non-negligible, it occurs exclusively during the incubation regime where absolute strain magnitudes remain minimal, thus having a negligible impact on the overall structural assessment. Furthermore, the final cumulative discrepancy of less than $2.5\%$ falls well within the conservative safety margins and regulatory uncertainties standard in nuclear core component design. Given that this localized error does not propagate but instead stabilizes asymptotically, the order-of-magnitude reduction in computational time provided by the accelerated IT-framework fully justifies its adoption as a high-fidelity surrogate for long-term operational cycles.

\begin{center}
    \nopagebreak
    \includegraphics[width=0.5\textwidth]{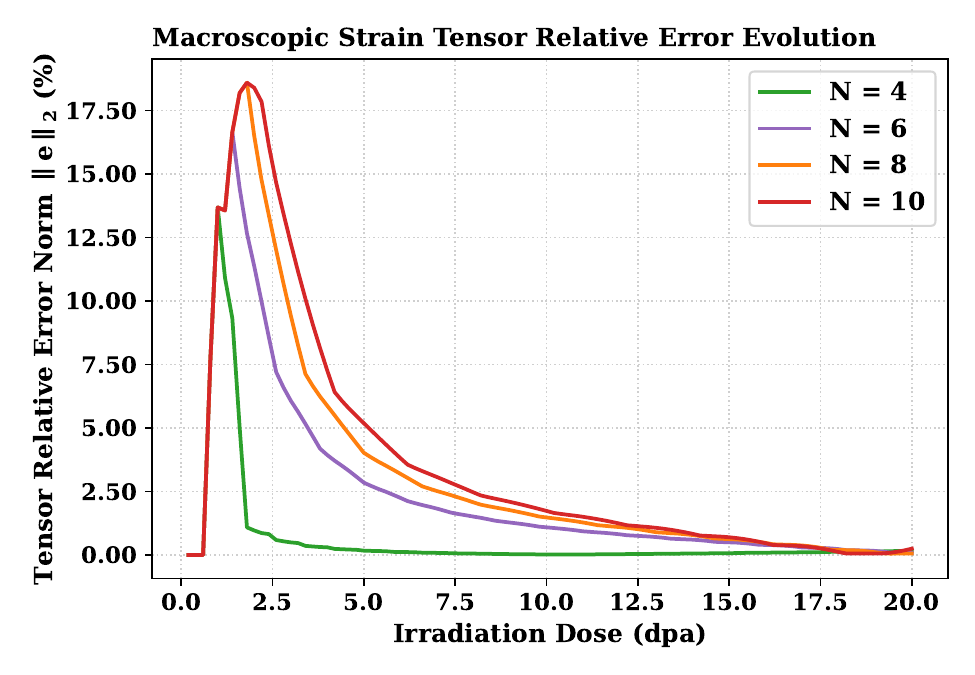}
    \captionof{figure}{Evolution of the macroscopic strain tensor relative error norm as a function of the irradiation dose for different linearization intervals $N$.}
    \label{fig:strain_error_evolution}
\end{center}
\FloatBarrier

Beyond the kinematic and constitutive accuracy, the primary advantage of the proposed sub-stepping formulation lies in its computational efficiency. Fig. \ref{fig:time_comparison} summarizes the total elapsed computational time required to complete the simulation for each studied configuration. While the standard direct-coupling scheme (Full VPSC) requires $1.11~\text{s}$ due to the execution of the crystal plasticity solver at every increment, introducing the interpolation table methodology leads to a reduction in CPU cost. For the $N = 4$ configuration, the computation time drops to $0.4209~\text{s}$, achieving a speedup of approximately $1.8\times$ while maintaining accuracy. As the refresh interval expands to $N = 6$ and $N = 8$, the CPU times further decrease to $0.3382~\text{s}$ and $0.3084~\text{s}$, respectively. For the $N = 10$ configuration, the total execution time is brought down to just $0.2497~\text{s}$, implying an overall reduction of nearly $72\%$ in the material-level computational effort (a $3.6\times$ speedup). These results suggest the practical utility of the linearization database approach, which balances microstructural tracking precision with the time-savings crucial for full-scale structural analysis.
\vspace{20mm}
\begin{center}
    \nopagebreak
    \includegraphics[width=0.5\textwidth]{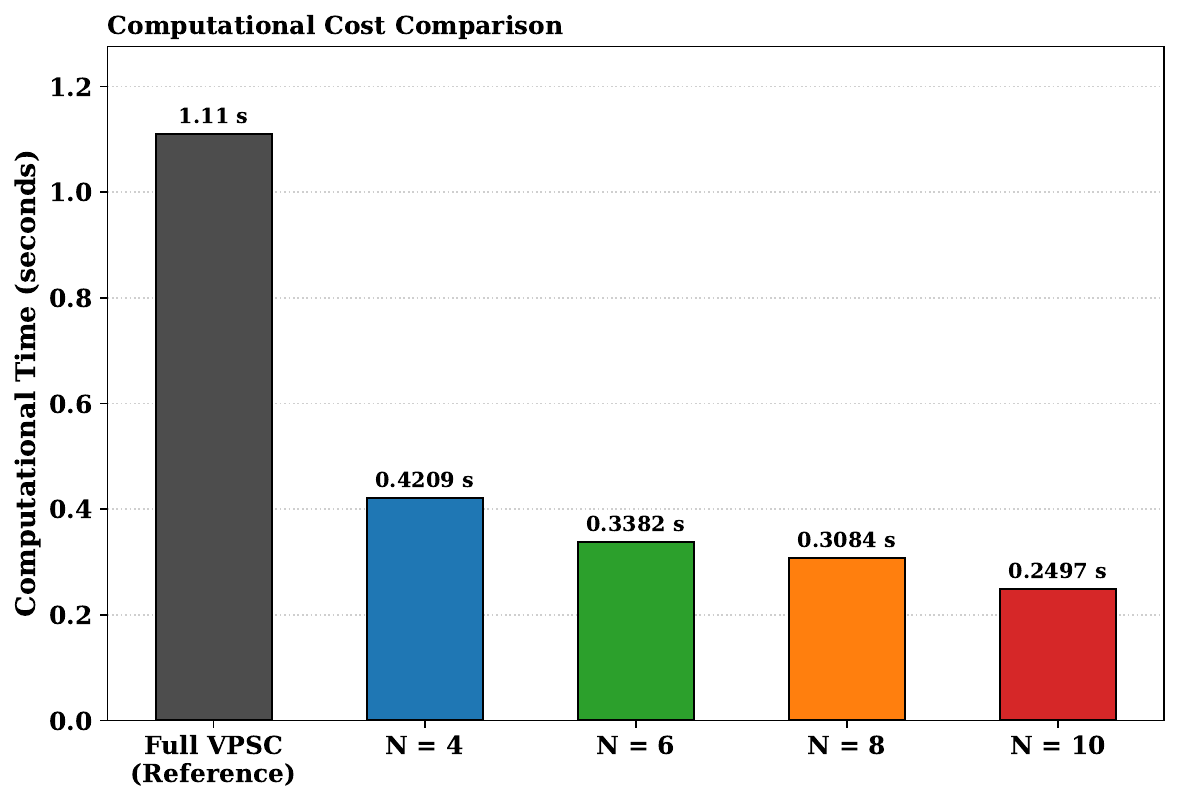}
    \captionof{figure}{Computational time comparison and overall efficiency gain between the standard concurrent direct-coupling VPSC simulation and the Taylor-expansion multi-scale interpolation scheme for different refresh intervals $N$. Labels above each bar denote the exact elapsed CPU time in seconds.}
    \label{fig:time_comparison}
\end{center}
\FloatBarrier
\subsection{Multi-scale Finite Element Application to Nuclear Components}
\label{subsec:structural_application}

To assess the capabilities of the VPSC-CAFEM interface via interpolation modeling at the component scale, a structural simulation was conducted using a representative nuclear spacer grid geometry in Code\_Aster. Specifically, the simulated domain consists of one half of
the central hexagon of a structural spacer grid, a critical core component subjected to severe irradiation conditions. The specific geometry and structural configuration of this grid component was previously reported and characterized by \citet{aguzzi2025toolbox}.

The initial material state was modeled using the reduced texture reported by \citet{patra2017finite}. Fig. \ref{fig:spacer_grid_geometry} schematically illustrates the isolated simulated domain of the spacer grid along with the basal pole figure representing its crystallographic orientation distribution function (ODF). In this context, RD, ND, and TD denote the rolling, normal, and transverse directions of the material reference frame, respectively. For physical context, the spatial relation with a adjacent fuel cladding tube is also displayed, although the tube was omitted from the finite element mesh.

\begin{center}
    \nopagebreak
    \includegraphics[width=0.35\textwidth]{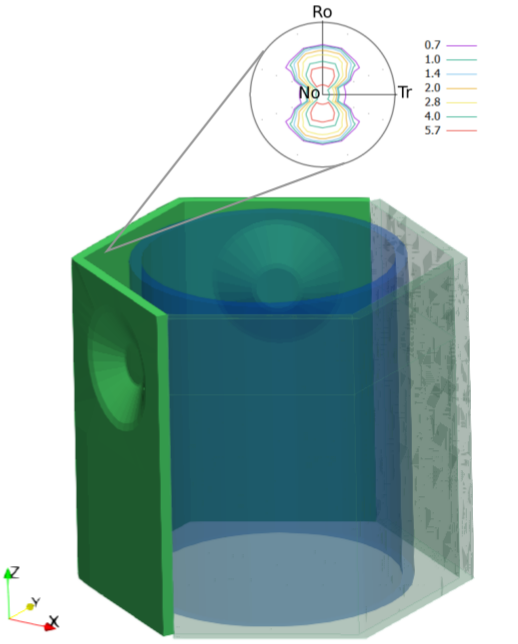}
    \captionof{figure}{Schematic of the simulated domain showing one-half of the central spacer grid hexagon adapted from \citet{aguzzi2025toolbox} and its spatial relation with an adjacent fuel cladding tube (omitted from the mesh). The inset displays the basal pole figure adapted from \citet{patra2017finite} used as the initial texture for VPSC calculations.}
    \label{fig:spacer_grid_geometry}
\end{center}

The component was exposed to a constant irradiation damage rate of $3.6 \times 10^{-4}$~dpa/h ($1.0 \times 10^{-7}$~dpa/s) over an extended operational cycle. The temporal discretization was defined by 50 fixed time increments of $\Delta t = 555.56$~\text{h} (with the time increment handled in hours within the solver), spanning a total simulated operational time of $27778$~h. 

Regarding the mechanical boundary conditions, realistic in-reactor constraints were introduced by imposing a fixed normal displacement of $1.0 \times 10^{-5}$~mm on the inner surfaces of the grid dimples. Restricting the analysis to the
spacer grid deformation allows the authors to evaluate the performance impact of the interpolation strategy without coupling the multi-scale framework with more complex contact non-linearities. Consequently, the localized constraint exerted by the expanding fuel cladding tube is directly emulated while avoiding the numerical cost associated with full mechanical contact algorithms. This approach was selected to isolate and evaluate the computational efficiency of the accelerated framework. By bypassing full contact-pair algorithms, the CPU time reduction achieved by the IT strategy can be accurately quantified. Consequently, the simulated domain is exclusively restricted to the grid piece, subjected solely to this mechanical displacement constraint alongside the uniform neutron flux distribution $\dot{\phi}$ that drives the irradiation phenomena.

\begin{figure*}[t!]
    \centering
% --- FILA SUPERIOR: DEFORMACIÓN EQUIVALENTE Y SU ERROR RELATIVO ---
    \begin{subfigure}[b]{0.32\textwidth}
        \centering
        \includegraphics[width=\textwidth]{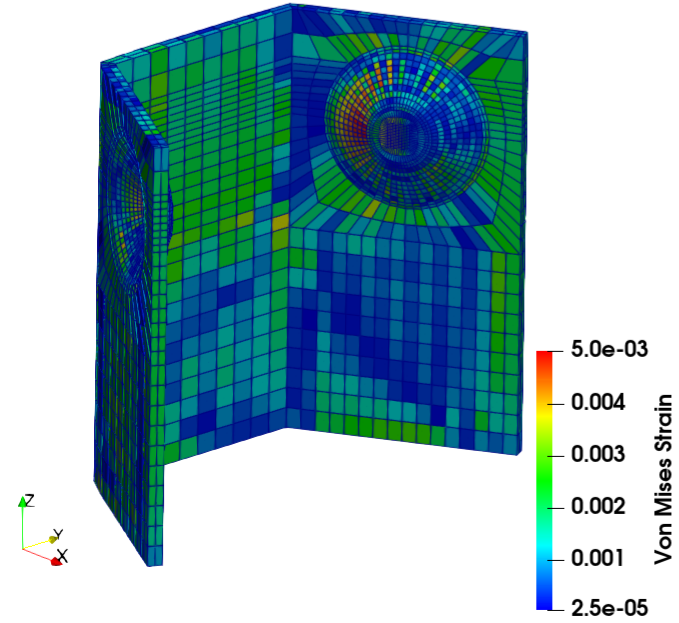}
        \caption{Full VPSC: Eq. strain}
        \label{fig:strain_fullvpsc}
    \end{subfigure}
    \hfill 
    \begin{subfigure}[b]{0.32\textwidth}
        \centering
        \includegraphics[width=\textwidth]{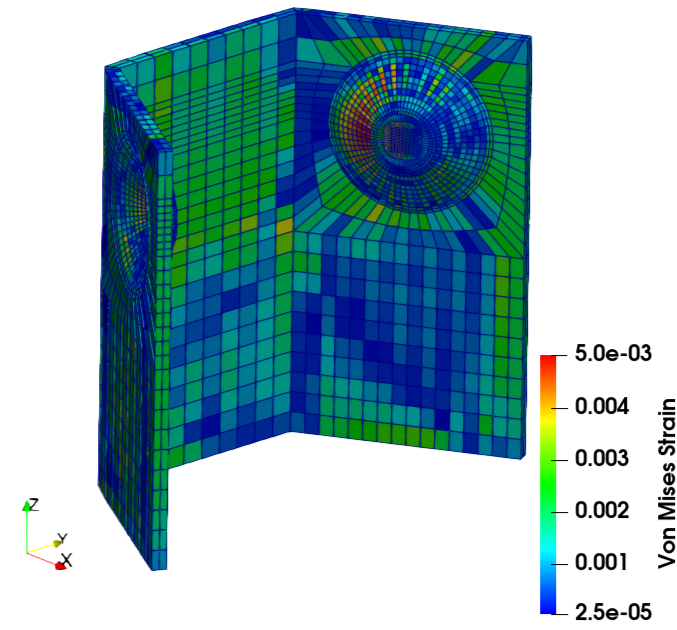}
        \caption{IT ($N=10$): Eq. strain}
        \label{fig:strain_N10}
    \end{subfigure}
    \hfill
    \begin{subfigure}[b]{0.32\textwidth}
        \centering
        \includegraphics[width=\textwidth]{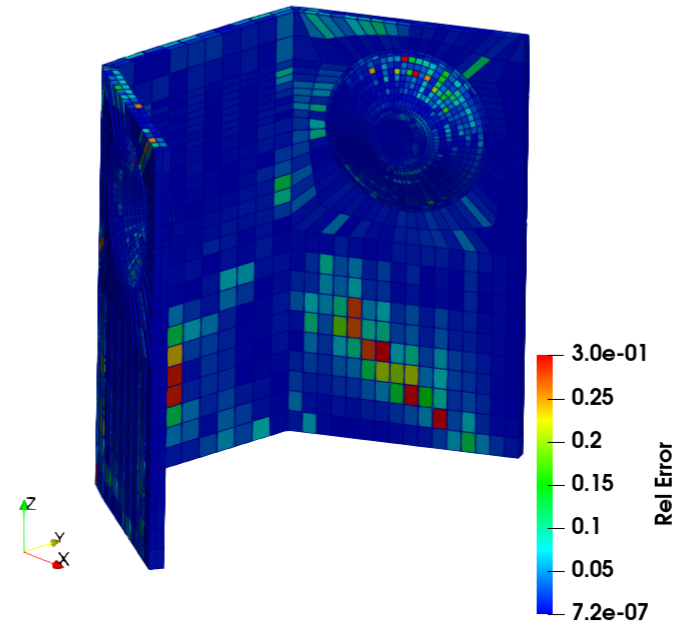}
        \caption{Strain Relative Error}
        \label{fig:strain_rel_error}
    \end{subfigure}
    
    \vspace{0.4cm} 

% --- FILA INFERIOR: TENSIÓN EQUIVALENTE Y SU ERROR RELATIVO ---
    \begin{subfigure}[b]{0.32\textwidth}
        \centering
        \includegraphics[width=\textwidth]{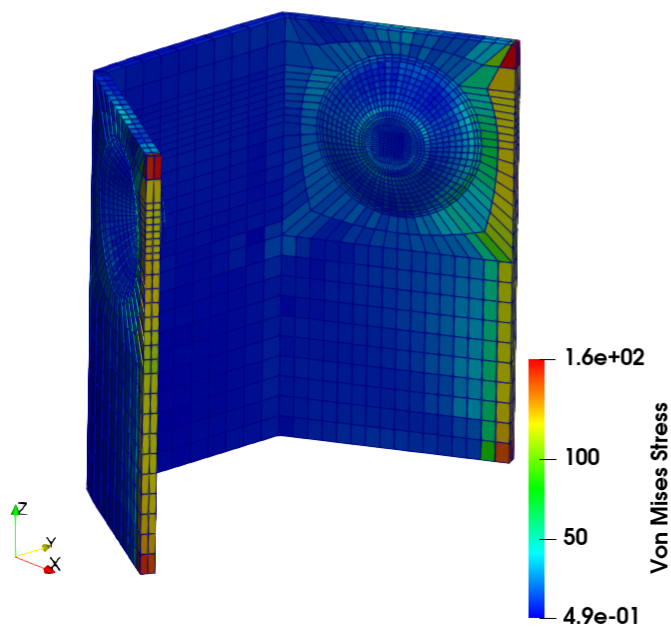}
        \caption{Full VPSC: Von Mises stress}
        \label{fig:stress_fullvpsc}
    \end{subfigure}
    \hfill 
    \begin{subfigure}[b]{0.32\textwidth}
        \centering
        \includegraphics[width=\textwidth]{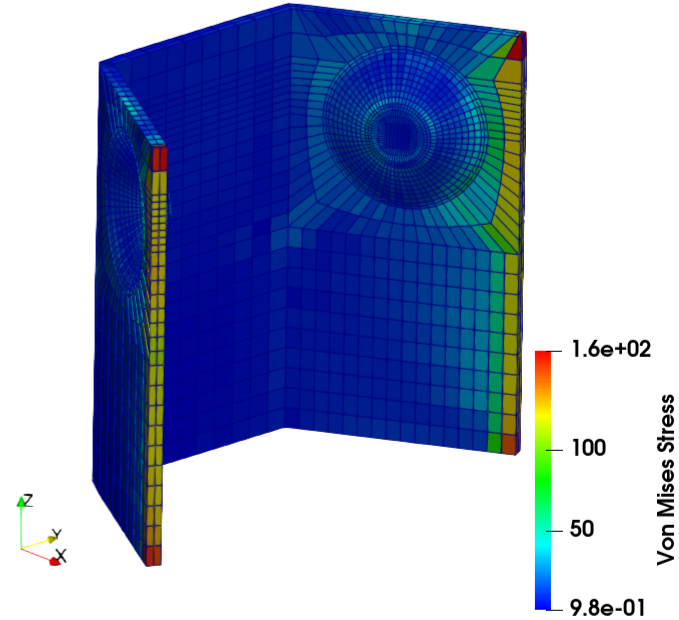}
        \caption{IT ($N=10$): Von Mises stress}
        \label{fig:stress_N10}
    \end{subfigure}
    \hfill
    \begin{subfigure}[b]{0.32\textwidth}
        \centering
        \includegraphics[width=\textwidth]{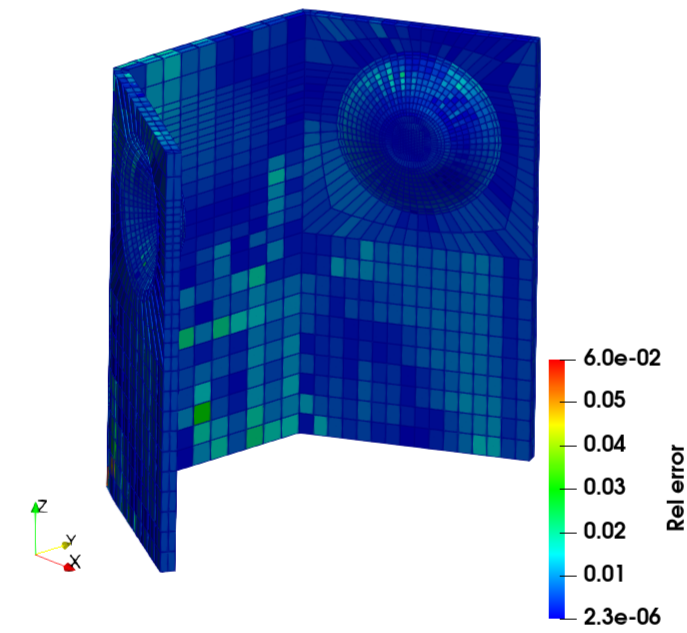}
        \caption{Stress Relative Error}
        \label{fig:stress_rel_error}
    \end{subfigure}

    \caption{Spatial distribution of macroscopic fields and relative errors in the simulated spacer grid component at the end of the irradiation cycle ($t = 27778$~h). Equivalent viscoplastic strain distribution for: (a) Direct Full VPSC, (b) Accelerated IT framework ($N=10$), and (c) associated strain relative error. Von Mises effective stress distribution for: (d) Direct Full VPSC, (e) Accelerated IT framework ($N=10$), and (f) associated stress relative error.}
    \label{fig:component_fields_comparison}
\end{figure*}

Fig. \ref{fig:component_fields_comparison} displays the spatial distribution of the equivalent accumulated viscoplastic strain ($\bar{\varepsilon}^{\text{vp}}$), the Von Mises effective stress ($\bar{\sigma}$), and their corresponding relative error fields at the end of the simulation.

% =============================================================================
% BLOQUE NUEVO: ECUACIONES DEL ERROR RELATIVO MACRO (TENSIÓN Y DEFORMACIÓN)
% =============================================================================
To quantify
the discrepancies shown in the third column of Fig. \ref{fig:component_fields_comparison}, pointwise relative errors at Gauss
points are evaluated. The scalar relative error metrics for the Von Mises effective stress ($e_{\sigma}$) and the equivalent accumulated viscoplastic strain ($e_{\varepsilon}$) are explicitly calculated as follows:
\begin{equation}
e_{\sigma} = \frac{|\bar{\sigma}_{\text{IT}} - \bar{\sigma}_{\text{ref}}|}{|\bar{\sigma}_{\text{ref}}|}
\end{equation}
\begin{equation}
e_{\varepsilon} = \frac{|\bar{\varepsilon}^{\text{vp}}_{\text{IT}} - \bar{\varepsilon}^{\text{vp}}_{\text{ref}}|}{|\bar{\varepsilon}^{\text{vp}}_{\text{ref}}|}
\end{equation}
where the subscripts $\text{Ref}$ and $\text{IT}$ denote the local fields computed via the reference fully concurrent Full VPSC solver and the accelerated Interpolation Table sub-stepping framework, respectively.

The structural response successfully captures the localized stress relaxation areas driven by irradiation creep, alongside the progressive dimensional changes induced by anisotropic growth directions under the high-flux environment. To evaluate the robustness of the accelerated framework under strict conditions, a point-by-point comparison is performed against the scenario corresponding to the largest linearization window ($N=10$). Even under this configuration, the accelerated IT framework reveals good spatial
agreement with the direct Full VPSC-FEM reference model across both macroscopic fields.

To assess the local accuracy in this configuration, the spatial distributions of the relative errors are analyzed in Figs. \ref{fig:strain_rel_error} and \ref{fig:stress_rel_error}. Regarding the effective stress field, the relative error remains low throughout the entire component, with a global maximum that does not exceed $6.0 \times 10^{-2}$ ($6\%$), and values well below $1\%$ across the majority of the finite element mesh. 

In the case of the equivalent viscoplastic strain field (Fig. \ref{fig:strain_rel_error}), the relative error is negligible over almost the entire domain. Although localized regions exhibiting apparent relative errors of $20-30\%$ are observed (indicated by the reddish elements near the dimple boundaries), a detailed inspection reveals that these hot-spots are localized strictly in zones where the absolute magnitude of the strain is close to zero. Consequently, even minute, nominal deviations in these stagnant regions yield artificially inflated relative values due to the vanishingly small denominator. Far from representing a mismatch in the calculation of the structural response, this behavior confirms that the error introduced by the Interpolation Table framework is negligible across all structurally critical regions of the component, preserving the essential microstructural physics of the polycrystal even when operating under the most unfavorable linearization conditions.

\subsection{Computational Efficiency and Speed-up Quantification}
\label{subsec:efficiency_quantification}

The primary advantage of the proposed PCYS interpolation strategy a significant
reduction in computational cost achieved during long-term simulations. The macro-scale implicit solver is forced to execute full, iterative VPSC cells at every single Newton-Raphson iteration for every integration point in the mesh. By replacing this computationally intensive local loop with a highly efficient 5D linear interpolation routine over the pre-computed PCYS database, and strictly restricting full micromechanical solver operations to the rate-calculation increments ($i_{\text{step}} \pmod N \in \{0, 1\}$), local execution times are minimized.

Table \ref{tab:computational_performance} summarizes the total micromechanical solver calls, the total continuum CPU time, the net acceleration factors, and the volume-averaged residual errors in the accumulated irradiation-induced strain tensor components ($\bar{\boldsymbol{\varepsilon}}^{\text{irr}}$) at the end of the operational transient.

\begin{table*}[t!]
    \small
    \centering
    \renewcommand{\arraystretch}{1.3}
    \caption{Computational performance metrics, net acceleration factors (Speed-up), and macroscopic kinematic error bounds of the accelerated IT-extrapolation framework compared against the fully concurrent direct VPSC-FEM reference scheme.}
    \label{tab:computational_performance}
    \resizebox{0.85\textwidth}{!}{%
        \begin{tabular}{l c c c c c}
            \toprule
            \textbf{Simulation Case} & 
            \textbf{\begin{tabular}[c]{@{}c@{}}Refresh\\ Interval ($N$)\end{tabular}} & 
            \textbf{\begin{tabular}[c]{@{}c@{}}Total VPSC\\ Calls\end{tabular}} & 
            \textbf{\begin{tabular}[c]{@{}c@{}}Total CPU\\ Time [s]\end{tabular}} & 
            \textbf{\begin{tabular}[c]{@{}c@{}}Net Speed-up\\ Factor\end{tabular}} & 
            \textbf{\begin{tabular}[c]{@{}c@{}}Final Integrated $\boldsymbol{\varepsilon}^{\text{irr}}$\\ Component Error [\%]\end{tabular}} \\
            \midrule
            Direct VPSC-FEM    & Reference & $1.45 \times 10^7$ & 1137.40 & $1.00\times$ (Ref.) & ---  \\
            IT + Extrapolation & $N = 4$    & $7.25 \times 10^6$ &  869.94 & $1.31\times$        & 1.06 \\
            IT + Extrapolation & $N = 6$    & $4.83 \times 10^6$ &  786.64 & $1.45\times$        & 0.24 \\
            IT + Extrapolation & $N = 8$    & $3.62 \times 10^6$ &  740.27 & $1.54\times$        & 0.32 \\
            IT + Extrapolation & $N = 10$   & $2.90 \times 10^6$ &  709.90 & $\mathbf{1.60\times}$ & 0.49 \\
            \bottomrule
        \end{tabular}%
    }
\end{table*}

The quantitative metrics demonstrate that the PCYS interpolation strategy yields a systematic, linear reduction in total VPSC calls as the refresh interval $N$ increases. This performance strictly adheres to the analytical $2/N$ operational ratio per sub-stepping block, which arises because the direct VPSC solver is invoked exactly twice every $N$ steps: first to establish the reference microstructural state ($\overline{M}_{\text{ref}}$) and immediately after to evaluate the tangent rate of change ($\overline{M}_{\text{check}}$). For the most accelerated configuration ($N=10$), the total number of VPSC calls drops drastically by $80\%$, falling from $1.45 \times 10^7$ down to $2.90 \times 10^6$.
\vspace{5mm}
\FloatBarrier
\begin{center}
    \centering
    \includegraphics[width=0.5\textwidth]{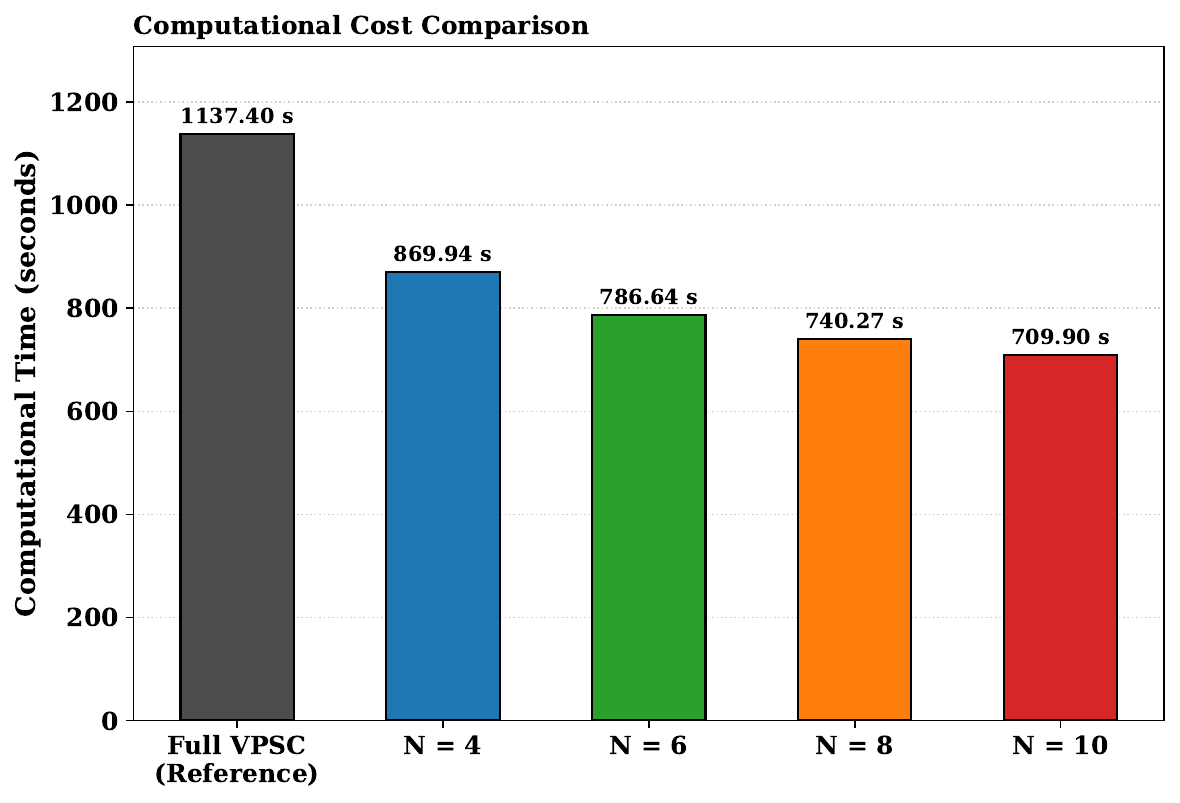}
    \captionof{figure}{Component-level computational cost comparison between the direct VPSC-FEM concurrent reference scheme and the accelerated IT-extrapolation sub-stepping strategy in Code\_Aster for different validation blocks $N$. Labels above each bar denote the exact execution time in seconds.}
    \label{fig:component_time_comparison}
\end{center}

As illustrated in Fig. \ref{fig:component_time_comparison}, this reduction in active micromechanical solver calls translates into a net component-level speed-up factor of $\mathbf{1.60\times}$, compressing the total macro-scale CPU time from $1137.40$~s down to $709.90$~s. 

A continuous tracking of the kinematic discrepancies reveals that the transient numerical drift is strictly bounded and phenomenologically driven. At the early stages of the irradiation regime ($t = 3333.4$~h), a localized peak error in the instantaneous strain rates is observed, reaching a maximum value of $9.45\%$ under the lowest sampling frequency ($N=10$). Rather than representing an algorithmic instability, this temporary drift is localized exclusively within the high-curvature region of the initial hardening and microstructural incubation curves, where the first-order Taylor expansion temporarily underpredicts the severe deceleration of the local inelastic rates. 

Crucially, as the underlying crystallographic structure approaches a steady-state regime at later stages ($t > 20000$~h), the self-correcting nature of the periodic VPSC recalibrations induces a remarkable asymptotic convergence. By the final operational increment ($t = 27778$~h), the cumulative component-specific error decays to negligible engineering values, dropping to $0.24\%$ for $N=6$ and maintaining a mere $0.49\%$ for $N=10$. This strong long-term accuracy confirms that the PCYS interpolation framework does not accumulate systemic drift over extended exposure periods, successfully preserving the underlying crystal plasticity physics while substantially expanding the numerical feasibility of large-scale structural analysis for core internals.
\section{Conclusions}\label{sec:conclusions}

In this work, a high-performance PCYS interpolation multi-scale framework has been developed and successfully integrated into the open-source FEM solver \texttt{Code\_Aster} to accelerate the structural simulation of nuclear components subject to coupled irradiation creep and growth. The proposed formulation demonstrates that the continuous microstructural drift, characterized by the viscoplastic compliance and back-extrapolated growth rate tensors, can be effectively decoupled from the instantaneous yield surface representation. While the use of an offline-generated, static Interpolation Table inherently assumes a stationary crystallographic texture and constant Critical Resolved Shear Stress values, the first-order Taylor extrapolation strategy successfully bridges this gap, capturing the history-dependent macroscopic response under high radiation flux without requiring computationally prohibitive on-the-fly table regenerations.

Furthermore, the local accuracy of employing micromechanical linearizations from the standalone viscoplastic self-consistent formulation was rigorously quantified through comparative benchmarks. When integrated into macro-scale finite element cases, this PCYS interpolation strategy demonstrates that a controlled, non-continuous application of the surrogate database introduces minimal numerical drift, keeping the resulting stress and strain fields well within strict engineering tolerances compared to fully concurrent $FE^2$ simulations. This architecture yields a reduction in computational cost, achieving acceleration factors that make long-term component-scale simulations highly viable by replacing thousands of iterative, nested micromechanical loops at each integration point with direct database queries and local first-order extrapolations.

Future work should address, several promising research avenues emerge to enhance the capabilities and robustness of the current multiscale framework. First, implementing more sophisticated multi-dimensional interpolation algorithms, such as Kriging or Gaussian Process Regression, could refine the mapping of the 5D Polycrystal Yield Surface, particularly in regions with high topological gradients or severe anisotropy. Second, the development of an explicit mathematical renormalization framework could dynamically correct the integrated historical error and numerical drift accumulated during the Taylor expansion blocks, ensuring long-term path-independent consistency. Finally, designing a physics-informed adaptive criteria to govern the refresh interval would allow the scheduling of full meso-scale VPSC calls to be triggered dynamically based on localized macro-scale indicators—such as sudden strain-path changes, stress triaxiality variations, or rapid accumulation of radiation damage—rather than adhering to a fixed step sequence.

\section*{CRediT authorship contribution statement}
F. E. Aguzzi: Methodology. Software, Formal analysis, Writing.  M. S. Armoa: Methodology, Writing. C. I Pairetti: Supervision, Writing - review \& editing. C. M. Venier: Supervision, Writing - review \& editing. A. E. Albanesi: Supervision, Writing - review \& editing.

\FloatBarrier

% ==========================================
% SECCIÓN DE NOMENCLATURA (Sin numerar como Apéndice)
% ==========================================
\FloatBarrier

\section*{Nomenclature} \label{sec:nomenclature}
\addcontentsline{toc}{section}{Nomenclature}

\subsection*{Latin Symbols}
\begin{description}[leftmargin=2.5cm, style=nextline, labelsep=0.5cm, font=\normalfont]
    \item[$b^j$] Normalized Burgers vector for the slip system $j$.
    \item[$B$] Crystallographic irradiation creep compliance.
    \item[$B_0$] Polycrystal compliance coefficient for isotropic approximation.
    \item[$C$] Fourth-order elastic stiffness tensor of the polycrystal.
    \item[$C^{\text{tg}}$] Consistent tangent operator tensor.
\item[$\mathbf{e}_{k}$] Macroscopic sample reference frame axes ($k \in \{x,y,z\}$).
    \item[$f$] Body force vector.
    \item[$i_{\text{step}}$] Current finite element simulation time step.
    \item[$J_{\text{NR}}$] Local Newton-Raphson Jacobian matrix.
    \item[$\overline{M}$] Polycrystal fourth-order viscoplastic compliance tensor.
    \item[$\overline{M}_{\text{tan}}$] Tangent polycrystal compliance tensor.
    \item[$\overline{M}_{\text{ref1}}$] Reference polycrystal compliance tensor evaluated at the recalibration step.
    \item[$\overline{M}_{\text{ref2}}$] Polycrystal compliance tensor evaluated at the step immediately following recalibration.
    \item[$\dot{\overline{M}}$] Numerical rate of change (slope) of the polycrystal viscoplastic compliance tensor.
    \item[$m_{kl}^j$] Symmetric Schmid tensor for the slip system $j$.
    \item[$N$] User-defined periodic VPSC recalibration interval.
    \item[$R$] Second-order rotation matrix.
    \item[$u$] Macroscopic displacement vector.
    \item[$X$] Non-linear local residual tensor field.
\end{description}

\subsection*{Greek Symbols}
\begin{description}[leftmargin=2.5cm, style=nextline, labelsep=0.5cm, font=\normalfont]
    \item[$\beta$] Internal state variable vector.
    \item[$\dot{\gamma}_{\text{creep}}^j$] Shear strain rate contributed by creep on system $j$.
    \item[$\Delta t$] Simulation time increment.
    \item[$\varepsilon_{\text{FE}}$] Total macroscopic strain tensor from solver.
    \item[$\dot{\varepsilon}_{kl}^{(\text{growth})}$] Microscopic irradiation growth strain-rate tensor.
    \item[$\dot{\varepsilon}_{kl}^{(\text{creep})}$] Microscopic irradiation creep strain-rate tensor.
    \item[$\dot{\bar{\varepsilon}}^{\text{vp}}$] Volume-averaged polycrystal viscoplastic strain rate.
    \item[$\bar{\varepsilon}^0$] Polycrystal back-extrapolated growth strain rate.
    \item[$\bar{\varepsilon}^0_{\text{tan}}$] Tangent polycrystal back-extrapolated growth strain rate.
    \item[$\bar{\dot{\varepsilon}}^0_{\text{ref1}}$] Reference polycrystal back-extrapolated growth strain rate evaluated at the recalibration step.
    \item[$\bar{\dot{\varepsilon}}^0_{\text{ref2}}$] Polycrystal back-extrapolated growth strain rate evaluated at the step immediately following recalibration.
    \item[$\ddot{\bar{\varepsilon}}^0$] Numerical rate of change (slope) of the polycrystal back-extrapolated growth strain rate tensor.
    \item[$\rho_d^j / \rho_{\text{ref}}$] Normalized dislocation density on system $j$.
    \item[$\sigma$] Macroscopic Cauchy stress tensor field.
    \item[$\bar{\sigma}$] Polycrystal volume-averaged stress tensor.
    \item[$\tau^j$] Resolved shear stress acting on slip system $j$.
    \item[$\phi$] Total accumulated radiation dose.
    \item[$\dot{\phi}$] Radiation dose rate.
    \item[$\chi$] Component-wise scalar error metric.
\end{description}

\subsection*{Subscripts, Superscripts and Operators}
\begin{description}[leftmargin=2.5cm, style=nextline, labelsep=0.5cm, font=\normalfont]
    \item[$(\cdot)^c$] Attribute corresponding to an individual grain $c$.
    \item[$(\cdot)^t, (\cdot)^{t+\Delta t}$] Variables at the beginning and end of the step.
    \item[$(\cdot)^^*$] Tensor mapped onto local material system.
    \item[$\langle \cdot \rangle$] Volumetric volume-average operator.
    \item[$\nabla_s$] Symmetric gradient operator.
\end{description}

\subsection*{Acronyms}
\begin{description}[leftmargin=2.5cm, style=nextline, labelsep=0.5cm]
    \item[CAFEM] Code Aster Finite Element Method.
    \item[CRSS] Critical Resolved Shear Stress.
    \item[dpa] Displacements per atom.
    \item[FEM] Finite Element Method.
    \item[HEM] Homogeneous Effective Medium.
    \item[IT] Interpolation Table.
    \item[ND] Normal Direction.
    \item[ODF] Orientation Distribution Function.
    \item[PCYS] Polycrystal Yield Surface.
    \item[RD] Rolling Direction.
    \item[RVE] Representative Volume Element.
    \item[TD] Transverse Direction.
    \item[VPSC] Viscoplastic Self-Consistent formulation.
\end{description}

\FloatBarrier
\appendix
% A partir de aquí el comando \appendix de cas-dc manejará las secciones como A, B, etc.

% ==========================================
% APÉNDICE A
% ==========================================
\section{Material parameters}
\label{apendiceA}
\gdef\theequation{A.\arabic{equation}} 
\gdef\thetable{A.\arabic{table}}       
\setcounter{equation}{0}
\setcounter{table}{0}

The material parameters presented in Table \ref{tab:zircaloy_params} were taken from \cite{patra2017crystal} and \cite{patra2017finite}.

\begin{table*}[h]
    \small
    \centering
    \caption{Model parameters for Zircaloy-2, from \cite{patra2017crystal} and \cite{patra2017finite}.}
    \renewcommand{\arraystretch}{1.2}
    \begin{tabular}{@{} p{0.15\textwidth} p{0.40\textwidth} p{0.45\textwidth} @{}} 
        \toprule
        \textbf{Parameter} & \textbf{Value} & \textbf{Meaning} \\
        \midrule
        $C_{ij}$ & \makecell[l]{$C_{11}, C_{22}, C_{33} = 143.5, 143.5, 164.9$ \\ $C_{12}, C_{13}, C_{23} = 72.5, 65.4, 65.4$ \\ $C_{44}, C_{55}, C_{66} = 32.1, 32.1, 35.5$} & Elastic constants in GPa for pure Zr, taken from \cite{simmons1965single} and \cite{kocks2000texture}. \\
        \midrule
        $f_r, f_{ic}$ & $0.97, 0.13$ & Fractions of point defect recombination and interstitial clustering. \\
        \midrule
        $B$ & $5.0 \times 10^{-5}$ MPa$\cdot$dpa$^{-1}$ & Crystallographic irradiation creep compliance. \\
        \midrule
        $\rho_{\text{ref}}$ & $2.26 \times 10^{14}$ m$^{-2}$ & Reference line dislocation density for the creep model. \\
        \midrule
        $b^j$ (pris.) & $3.0 \times 10^{-10}$ m & Burgers vector magnitude along prismatic directions ($j = \alpha_1, \alpha_2, \alpha_3$). \\
        \midrule
        $b^j$ (basal) & $5.0 \times 10^{-10}$ m & Burgers vector magnitude along the basal direction ($j = c$). \\
        \bottomrule
    \end{tabular}
    \label{tab:zircaloy_params}
\end{table*}

% ==========================================
% APÉNDICE B
% ==========================================
\section{Generation of the Interpolation Table (IT)}
\label{apendiceB}
\gdef\theequation{B.\arabic{equation}} 
\setcounter{equation}{0}

The Interpolation Table (IT) is constructed by probing the Polycrystal Yield Surface (PCYS) in a 5D deviatoric space. Following the formulation described in \cite{tome2023material}, any deviatoric state (stress or strain rate) can be represented as a vector in this space, whose orientation is defined by four angular coordinates: $\theta_1, \theta_2, \theta_3,$ and $\theta_4$. The mathematical basis of the 5D angular discretization, probing strategy, and optimization routines used to construct this database are detailed in this section.

\subsection{Angular Discretization and Probing}

The algorithm generates a grid of equispaced unit vectors. For each state, the components of the normalized "action" vector $\mathbf{a}$ (imposed strain rate in this work) are computed as:
\begin{equation}
\begin{aligned}
a_1 &= \sin\theta_1 \sin\theta_2 \sin\theta_3 \sin\theta_4 \\
a_2 &= \cos\theta_1 \sin\theta_2 \sin\theta_3 \sin\theta_4 \\
a_3 &= \cos\theta_2 \sin\theta_3 \sin\theta_4 \\
a_4 &= \cos\theta_3 \sin\theta_4 \\
a_5 &= \cos\theta_4
\end{aligned}
\end{equation}

The domain is partitioned using the parameter $N_{\text{part}}$. The angular step is defined as $\Delta\theta = \pi / (2 N_{\text{part}})$. In this study, $N_{\text{part}}=4$ was adopted, providing a balanced angular resolution of $22.5^\circ$. For non-centro-symmetric properties, the scan for $\theta_1$ covers the range $[-\pi, \pi]$, while for centro-symmetric cases, the range $[0, \pi]$ is sufficient.

\subsection{Optimization, Redundancy Handling, and Local Interpolation}

To minimize the number of calls to the VPSC solver during the table generation, the algorithm identifies geometric singularities in the 5D spherical representation. When a higher-order angle is zero, the probe becomes independent of the lower-order angles. The implemented code detects these redundant states to skip unnecessary calculations:
\begin{itemize}
    \item If $\theta_4 = 0$, the vector components are $(0, 0, 0, 0, 1)$ regardless of $\theta_1, \theta_2, \theta_3$.
    \item If $\theta_3 = 0$ (and $\theta_4 \neq 0$), the vector is independent of $\theta_1, \theta_2$.
    \item If $\theta_2 = 0$ (and $\theta_3, \theta_4 \neq 0$), the vector is independent of $\theta_1$.
\end{itemize}

For each unique probe resulting from this optimized scan, the VPSC model computes the corresponding "reaction" vector $\mathbf{r}$ (macroscopic stress response). The resulting pairs $\{\mathbf{a}, \mathbf{r}\}$ are stored in an ordered binary database.

During the macro-scale FEM structural analysis, when Code\_Aster requests a material response for an arbitrary input state $\mathbf{x}$ lying within the 5D deviatoric domain, a local multi-linear interpolation routine is triggered. The algorithm first identifies the $2^5 = 32$ bounding grid vertices (the enclosing hypercube) surrounding $\mathbf{x}$. The interpolated response $\mathbf{y}(\mathbf{x})$ is then evaluated via a weighted combination of the pre-calculated reactions $\mathbf{y}_k$ at each vertex:
\begin{equation}\label{eq:5D_interpolation}
    \mathbf{y}(\mathbf{x}) = \sum_{k=1}^{32} \omega_k \, \mathbf{y}_k
\end{equation}
where the scalar weighting factors $\omega_k$ are calculated as products of the normalized distances between $\mathbf{x}$ and the respective hypercube facets:
\begin{equation}\label{eq:weights_interpolation}
    \omega_k = \prod_{d=1}^{5} \left(1 - \frac{|x_d - x_{k,d}|}{\Delta \theta_d}\right)
\end{equation}
In Eq. \ref{eq:weights_interpolation}, $x_d$ represents the coordinate of the evaluation point along the $d$-th dimension, $x_{k,d}$ is the coordinate of vertex $k$, and $\Delta \theta_d$ denotes the localized grid spacing. This efficient algebraic evaluation allows Code\_Aster to bypass the costly iterative polycrystal equations during the local macro-solver steps, securing the computational speed-ups discussed in Section \ref{sec:resultados}.
\FloatBarrier
%% Bibliografía (basada en el estilo natbib del original )
\bibliographystyle{cas-model2-names}
\FloatBarrier

\end{document}